\documentclass[prl,aps,showpacs,groupedaddress,superscriptaddress,twocolumn,toc=flat]{revtex4-2}

\usepackage{physics}
\usepackage{amsmath,amssymb}
\usepackage{graphicx}
\usepackage[dvipsnames]{xcolor}
\usepackage[caption=false]{subfig}
\usepackage[colorlinks=true,linkcolor=blue,urlcolor=blue,citecolor=blue]{hyperref}

\usepackage{todonotes}

\begin{document}
\title{Bosonic Pfaffian State in the Hofstadter-Bose-Hubbard Model}

\author{F. A. Palm}
\affiliation{Department of Physics and Arnold Sommerfeld Center for Theoretical Physics (ASC), Ludwig-Maximilians-Universit\"at M\"unchen, Theresienstr. 37, D-80333 M\"unchen, Germany}
\affiliation{Munich Center for Quantum Science and Technology (MCQST), Schellingstr. 4, D-80799 M\"unchen, Germany}

\author{M. Buser}
\affiliation{Department of Physics and Arnold Sommerfeld Center for Theoretical Physics (ASC), Ludwig-Maximilians-Universit\"at M\"unchen, Theresienstr. 37, D-80333 M\"unchen, Germany}
\affiliation{Munich Center for Quantum Science and Technology (MCQST), Schellingstr. 4, D-80799 M\"unchen, Germany}

\author{J. L\'eonard}
\affiliation{Department of Physics, Harvard University, Cambridge, Massachusetts 02138, USA}

\author{M. Aidelsburger}
\affiliation{Munich Center for Quantum Science and Technology (MCQST), Schellingstr. 4, D-80799 M\"unchen, Germany}
\affiliation{Department of Physics, Ludwig-Maximilians-Universit\"at M\"unchen, Schellingstr. 4, D-80799 M\"unchen, Germany}

\author{U. Schollw\"ock}
\affiliation{Department of Physics and Arnold Sommerfeld Center for Theoretical Physics (ASC), Ludwig-Maximilians-Universit\"at M\"unchen, Theresienstr. 37, D-80333 M\"unchen, Germany}
\affiliation{Munich Center for Quantum Science and Technology (MCQST), Schellingstr. 4, D-80799 M\"unchen, Germany}

\author{F. Grusdt}
\affiliation{Department of Physics and Arnold Sommerfeld Center for Theoretical Physics (ASC), Ludwig-Maximilians-Universit\"at M\"unchen, Theresienstr. 37, D-80333 M\"unchen, Germany}
\affiliation{Munich Center for Quantum Science and Technology (MCQST), Schellingstr. 4, D-80799 M\"unchen, Germany}

\date{\today}

\begin{abstract}
Topological states of matter, such as fractional quantum Hall states, are an active field of research due to their exotic excitations. In particular, ultracold atoms in optical lattices provide a highly controllable and adaptable platform to study such new types of quantum matter. However, finding a clear route to realize non-Abelian quantum Hall states in these systems remains challenging. Here we use the density-matrix renormalization-group (DMRG) method to study the Hofstadter-Bose-Hubbard model at filling factor $\nu = 1$ and find strong indications that at $\alpha=1/6$ magnetic flux quanta per plaquette the ground state is  a lattice analog of the continuum non-Abelian Pfaffian. We study the on-site correlations of the ground state, which indicate its paired nature at $\nu = 1$, and find an incompressible state characterized by a charge gap in the bulk. We argue that the emergence of a charge density wave on thin cylinders and the behavior of the two- and three-particle correlation functions at short distances provide evidence for the state being closely related to the continuum Pfaffian. The signatures discussed in this letter are accessible in current cold atom experiments and we show that the Pfaffian-like state is readily realizable in few-body systems using adiabatic preparation schemes.
\end{abstract}

\maketitle

\textit{Introduction.---} During the past decades it was realized that the interplay of interactions and topology gives rise to exotic phases of matter, exhibiting features like quantum number fractionalization~\cite{laughlin_83} or excitations obeying fractional braiding statistics, which evade the classification in bosons and fermions~\cite{halperin_84, arovas_84}. One of the first microscopic states found to possess \mbox{non-Abelian} braiding statistics was Moore~and~Read's Pfaffian~\cite{moore_91}. There has been a long and still ongoing debate whether electronic fractional quantum Hall (FQH) systems at filling $\nu = 5/2$ may realize this exotic paired state of matter~\cite{haldane_88, morf_98, rezayi_00, willett_13, son_15, sun_20}.

The direct demonstration of non-Abelian braiding~\cite{park_16, dai_17, song_18} in extended systems~\cite{bonderson_06} has become one of the biggest challenges of modern experimental physics. Quantum simulators with ultracold atoms offer a promising experimental platform where interferometric probes of topological invariants have already been demonstrated~\cite{atala_13, duca_15} and extensions of such methods to anyons are possible~\cite{grusdt_16, nakamura_20}. In recent years, significant advances regarding ultracold atoms in optical lattices have led to the implementation of Hofstadter Hamiltonians~\cite{aidelsburger_13, miyake_13}, even in the presence of interactions~\cite{tai_17}, and the realization of topological states of matter in two-dimensional (2D) systems in general~\cite{jotzu_14, flaeschner_16, goldman_16, aidelsburger_18}. However, any attempt to detect anyonic excitations first requires the identification and realization of a suitable Hamiltonian as well as preparation schemes to reach the desired ground state.

Here we argue that the Pfaffian is readily realizable in cold atom experiments implementing the Hofstadter-Bose-Hubbard (HBH) model~\cite{tai_17}. To this end, we perform density-matrix renormalization-group (DMRG) studies on extended cylinders and demonstrate that the ground state of the HBH model on a square lattice at filling factor $\nu = 1$ is a lattice analog of the Pfaffian. We find a significant bulk gap, which allows for an experimental realization of the Pfaffian in current ultracold atomic systems. Strongly suppressed three-particle correlations at short distances corresponding to screened three-particle interactions provide a directly accessible signature for the paired nature of the ground state. On thin cylinders the topologically ordered Pfaffian evolves into a charge density wave (CDW) from which adiabatic preparation of the Pfaffian should be possible in extended systems. For small systems with a few bosons, we propose a direct adiabatic pathway into the Pfaffian.

Earlier attempts to study FQH physics in cold atoms used rotating Bose-Einstein condensates to mimic an effective magnetic field~\cite{abo-shaeer_01, schweikhard_04}. Reaching the quantum degenerate regime turned out to be challenging, although signatures of a bosonic Laughlin state at $\nu = 1/2$ have been observed in rotating microtraps~\cite{gemelke_10}. In addition, numerical studies involving pure contact interactions in the lowest Landau level found the bosonic Pfaffian at filling $\nu=1$~\cite{regnault_03, regnault_04}. Here we study the same physics in a cylindrical lattice system relatively close to the continuum limit. The cylinder geometry, even in the continuum, places the FQH states close to CDW states in the quasi-one-dimensional limit, while the topologically ordered FQH states are restored in the 2D limit~\cite{rezayi_94, bergholtz_06, seidel_06}. Exact diagonalization studies of the HBH model on small, toroidal systems have found indications for a lattice analog of the Pfaffian~\cite{sterdyniak_12}. However, the robustness of the state with respect to strong two-particle interactions in larger systems remained unclear. Lattice effects similar to those discussed here have previously been studied at filling $\nu = 1/2$~\cite{sorensen_05, hafezi_07} and $\nu = 2$~\cite{he_17}. Related lattice versions of the Pfaffian were found in spin systems~\cite{greiter_09}, which led to the proposal of a parent Hamiltonian~\cite{greiter_14}, and in the Haldane model with three-body on-site interactions~\cite{wang_12}.

\begin{figure}
	\includegraphics[width=\linewidth]{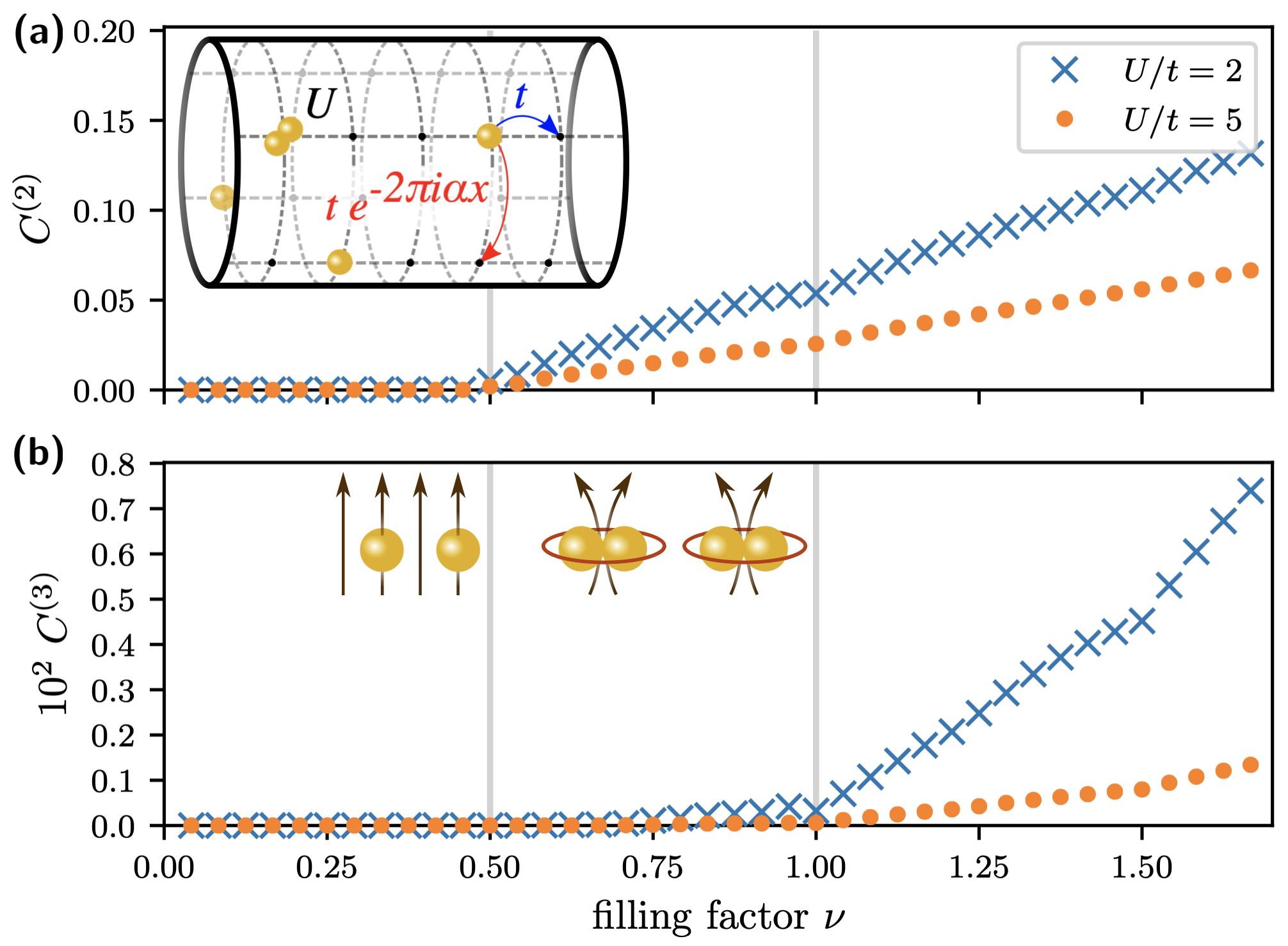}
	\caption{(a) Two- and (b) three-particle on-site correlations per particle as a function of the filling factor $\nu$ for fixed flux $\alpha=1/6$, different two-particle interaction strengths $U$, at most $N_{\rm max}=3$ particles per site, $L_y = 4$, and $L_x = 37$. The sudden appearance of correlations at $\nu = 1/2$ and $1$ indicate the existence of the Laughlin and Pfaffian state, respectively. In the corresponding regions, the screened composite particles are illustrated. The inset in (a) illustrates the model.
	\label{fig:InteractionEnergies}}
\end{figure}

\textit{Model.---} We study the Bose-Hubbard model with the usual two-body contact interactions on an $L_x\times L_y$-square lattice with lattice constant $a$ assuming periodic boundary conditions in the short $y$-direction, thus realizing the square lattice on a thin cylinder (see inset in Fig.~\ref{fig:InteractionEnergies}(a)). The lattice is subject to a perpendicular magnetic field with flux $\alpha$ per plaquette in units of the magnetic flux quantum. The resulting HBH Hamiltonian in the Landau gauge reads\pagebreak
\begin{widetext}
\begin{equation}
\hat{\mathcal{H}} = -t \sum_{y=1}^{L_y}\sum_{x=1}^{L_x-1} \left(\hat{a}^{\dagger}_{x+1, y}\hat{a}^{\vphantom{\dagger}}_{x,y} + \mathrm{H.c.}\right) -t \sum_{y=1}^{L_y}\sum_{x=1}^{L_x} \left(\mathrm{e}^{2\pi i \alpha x}\hat{a}^{\dagger}_{x,y+1}\hat{a}^{\vphantom{\dagger}}_{x,y} + \mathrm{H.c.}\right) +\frac{U}{2}\sum_{x,y} \hat{n}_{x,y}\left(\hat{n}_{x,y}-1\right),
\label{Eq:HBH-Hamiltonian}
\end{equation}
\end{widetext}
where $\hat{a}^{(\dagger)}_{x,y}$ annihilates (creates) a boson at site \mbox{$\mathbf{i} = (x, y \mod L_y)$} and $\hat{n}_{x,y} = \hat{a}_{x,y}^{\dagger}\hat{a}^{\vphantom{\dagger}}_{x,y}$ is the boson number operator. The first (second) term of the Hamiltonian describes hopping with amplitude $t$ between neighboring sites along $x$ ($y$), the last term describes repulsive ($U/t > 0$) on-site interactions.

For periodic boundary conditions in the $y$-direction the flux per plaquette $\alpha$ and the total number of flux quanta $N_{\phi}$ are related by \mbox{$\alpha = N_{\phi}/[L_y \left(L_x - 1\right)]$}.
In this letter, we restrict ourselves to the case $\alpha = 1/6$ and dilute systems of $N$ bosons, $N \ll L_xL_y$, close to the continuum limit to be able to relate our findings to earlier results for bosonic FQH systems~\cite{sorensen_05}. Furthermore, we focus on the regime close to the filling factor $\nu = N/N_{\phi} = 1$.

\textit{Screened interactions.---} In the continuum limit of Eq.~\eqref{Eq:HBH-Hamiltonian} at $\nu=1$ the ground state is well described by the Pfaffian~\cite{cooper_01}, which is the exact zero-energy ground state of the repulsive three-body parent Hamiltonian~\cite{greiter_91}
\begin{equation}
\hat{\mathcal{H}}^{(3)} \propto \sum_{i,j,k} \delta(z_i-z_j) \delta(z_j - z_k).
\label{Eq:parentHamiltonian}
\end{equation}
This Hamiltonian allows for two particles at the same location but penalizes three particles at the same point in space. Similar to the way the $\nu = 1/2$ Laughlin state fully screens local two-body interactions, the $\nu = 1$ Pfaffian has vanishing energy with respect to the parent Hamiltonian~\eqref{Eq:parentHamiltonian}~\cite{jain_89, cooper_99}. This can be understood in the composite fermion (CF) picture, where some amount of flux is attached to each constituent boson, such that the combined object has fermionic statistics. The $\nu = 1/2$ Laughlin state can be described using non-interacting CFs with one flux quantum attached to each boson. In the $\nu = 1$ Pfaffian the same CFs form pairs which experience a screened interaction among each other~\cite{scarola_00}, thus resulting in a vanishing three-boson interaction energy.

Using the single-site variant~\cite{hubig_15} of the DMRG method~\cite{white_92, schollwoeck_11}, we calculate the canonical ground state of the HBH Hamiltonian $\hat{\mathcal{H}}$, allowing for at most $N_{\rm max}=3$ bosons per site. We consider a finite system \mbox{$(L_x=37, L_y = 4)$} and vary the two-particle interaction strength, $U/t = 2, 5$.
We calculate the ground state's two- and three-particle on-site correlations per particle,
\begin{align}
C^{(2)} &= \sum_{\mathbf{i}} \left\langle \hat{n}_{\mathbf{i}}\left(\hat{n}_{\mathbf{i}} - 1\right) \right\rangle/N,\\
C^{(3)} &= \sum_{\mathbf{i}} \left\langle \hat{n}_{\mathbf{i}} \left(\hat{n}_{\mathbf{i}} - 1\right) \left(\hat{n}_{\mathbf{i}} - 2\right) \right\rangle/N,
\end{align}
for a broad range of filling factors.

We show in Fig.~\ref{fig:InteractionEnergies}(a) that the two-particle on-site correlation essentially vanishes below $\nu = 1/2$, a key signature for the $1/2$ Laughlin state. Similarly, in Fig.~\ref{fig:InteractionEnergies}(b) we show that the three-particle on-site correlation is strongly suppressed for filling factors up to $\nu = 1$, a key signature for the Pfaffian related to the screened three-particle interactions in its parent Hamiltonian.

\begin{figure}
	\includegraphics[width=\linewidth]{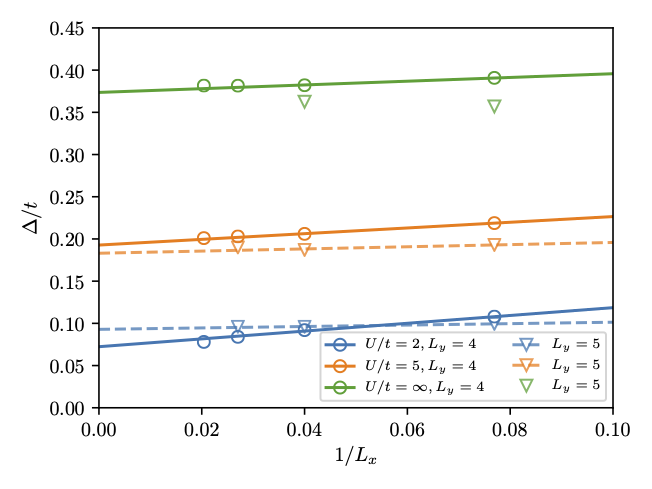}
	\caption{\label{fig:FiniteSizeScaling}Charge gap $\Delta$ for different system sizes $(L_x, L_y)$ and interaction strengths $U/t$ at $\nu = 1$ with at most $N_{\rm max}=3$ bosons per site. The size of the gap on infinite cylinders is estimated from a linear fit and depends on both the interaction strength $U/t$ and the circumference $L_y$ of the cylinder.}
\end{figure}
\begin{table}
	\begin{tabular}{|c||c|c||c|c|}
		\hline
		$U/t$ & \multicolumn{2}{c||}{$\Delta/t$} & \multicolumn{2}{c|}{$n_1$} \\
		\hline
		& $L_y = 4$ & $L_y = 5$ & $L_y = 4$ & $L_y = 5$\\
		\hline\hline
		$2$ & $0.072(3)$ & $0.093(5)$ & $0.055(2)$ & $0.051(1)$\\
		\hline
		$5$ & $0.193(2)$ & $0.183(9)$ & $0.0528(7)$ & $0.0420(7)$\\
		\hline
		$\infty$ & $0.3736(6)$ & --- & $0.0382(5)$ & --- \\
		\hline
	\end{tabular}
	\caption{\label{tab:FiniteSizeScaling}Extrapolated charge gap $\Delta/t$ and CDW order parameter $n_1$ for an infinite cylinder as obtained by a linear fit of the finite-size results shown in Figs.~\ref{fig:FiniteSizeScaling}~and~\ref{fig:ScalingCDW}. For a discussion of the errors see~\cite{supp}.}
\end{table}

\textit{Charge gap and incompressibility.---} While the screened two- and three-particle interactions at $\nu = 1$ provide an experimentally accessible indicator for the presence of a ground state related to the Pfaffian, they do not reveal further insight into the nature of the state.

The continuum Pfaffian is an incompressible state with a charge gap in the bulk~\cite{read_99}. To investigate this property in the lattice systems from Eq.~\eqref{Eq:HBH-Hamiltonian}, we determine the
%canonical
ground state of the HBH model for different parameters $U/t = 2, 5, \infty$, \mbox{$L_x = 13, 25, 37, 49,$} and \mbox{$L_y = 4, 5$}.
%Given the small remaining three-particle correlations at $\nu = 1$, Fig.~\ref{fig:InteractionEnergies}(b), and the low particle number density, we restrict the local occupation to at most $N_{\rm max}=2$ bosons per site from now on.

We find a charge gap at $\nu = 1$ indicated by a plateau in $\nu(\mu)$
%, i.e. $\frac{\partial\nu}{\partial\mu} = 0$ over a finite range of $\mu$,
as expected for an incompressible phase~\cite{supp}. In Fig.~\ref{fig:FiniteSizeScaling}, we show that strong interactions stabilize the incompressible phase and result in a large gap, which increases with the repulsion strength. For $U/t = 2, 5$, the obtained gap extrapolates well to the thermodynamic limit, $1/L_x \to 0$, and the extrapolated charge gaps are given in Tab.~\ref{tab:FiniteSizeScaling}. Because the circumference $L_y = 4, 5$ is comparatively short, on the order of the magnetic length, gapless edge states at the ends of the cylinder cannot be resolved yet. For hard-core bosons, $U/t = \infty$, and $L_y = 5$ numerical convergence is difficult and the nature of the ground state in the thermodynamic limit remains unclear.

\begin{figure}[t]
	\includegraphics[width=\linewidth]{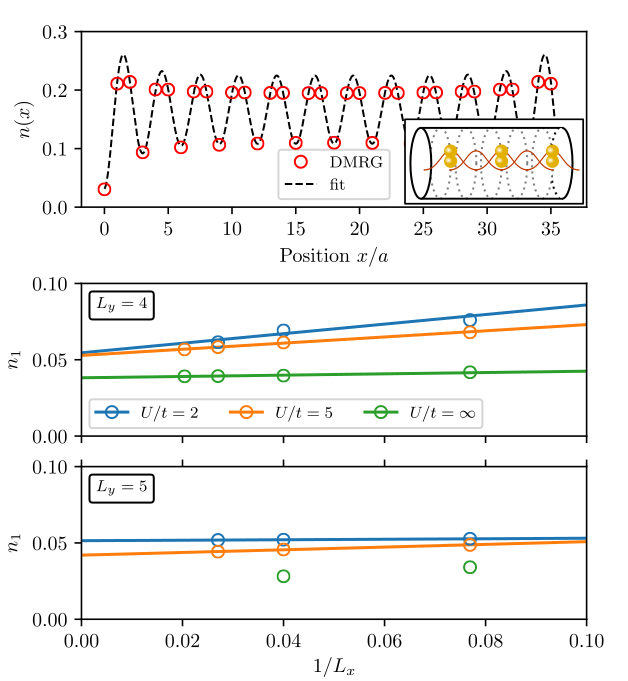}
	\caption{\label{fig:ScalingCDW} (a) Density profile $n(x)$ of the ground state for \mbox{$L_x=37, L_y = 4, U/t = 5, N_{\rm max}=3$} at $\nu = 1$ ($N=24$) showing a clear CDW. The maxima correspond to the occupied Landau level orbitals of the $\ket{\hdots2020\hdots}$ Tao-Thouless state (inset). We fitted the function in Eq.~\eqref{Eq:FitFunctionCDW} to the numerical data using all parameters. (b,c) CDW order parameter for $L_y = 4$ (b) and $L_y=5$ (c) with at most $N_{\rm max}=3$ bosons per site. The limit of infinite cylinders is extrapolated by a linear fit at $\nu=1$. As compared to $L_y=4$, for $L_y = 5$ the order parameter is
	%significantly
	smaller at all interaction strengths.}
\end{figure}

\textit{CDW on thin cylinders.---} The average density
\begin{equation}
n(x)=\sum_{y}\left\langle\hat{n}_{x,y}\right\rangle/L_y
\end{equation}
reveals a pronounced charge density wave (CDW) for all considered parameters, see Fig.~\ref{fig:ScalingCDW}(a). The rapid decay of enhanced density modulations at the edges shows that this is indeed a bulk property and not an edge effect. While the 2D FQH system is topologically ordered, we attribute the symmetry-breaking ground state found here to the finite system size and in particular the thin cylinders. Indeed, for continuous FQH systems on cylinders the Laughlin states in the 2D limit ($L_y \to \infty$) are adiabatically connected to symmetry-breaking Tao-Thouless states in the limit of thin cylinders ($L_y \to 0$)~\cite{rezayi_94}. For the bosonic case at $\nu =1$ studied here, the interplay of the CDW and the Pfaffian was discussed in the continuum by Seidel et al.~\cite{seidel_06} and Bergholtz et al.~\cite{bergholtz_06}.

To extract the CDW order parameter, we fit the density distribution by
\begin{equation}
n(x) = n_0 + n_1\sin(kx+\phi_0)\left(1+\eta(x, A, \xi)\right),\label{Eq:FitFunctionCDW}
\end{equation}
with $\eta(x, A, \xi) = A/2(\exp[-x/\xi] + \exp[-(L_x-1-x)/\xi])$ capturing the decay at the edges. For a system of $N$ particles the CDW has $N/2$ maxima corresponding to the occupied Landau level orbitals of the $\ket{\hdots2020\hdots}$ Tao-Thouless state in the limit $L_y\to0$.
%For $U/t = 0$ edge effects tend to affect the structure of the CDW more drastically\cite{supp}.
We extract the CDW order parameter $n_1$ from the fit and extrapolate the finite size results to the limit $L_x \to \infty$ as shown in Fig.~\ref{fig:ScalingCDW}(b,c). The extrapolated order parameter for the $L_y =4$ cylinder is
%substantially
larger than for the wider $L_y = 5$ cylinder, independently of the interaction strength, in agreement with the continuum results~\cite{seidel_06, bergholtz_06, supp}.

\begin{figure}
	\includegraphics[width=\linewidth]{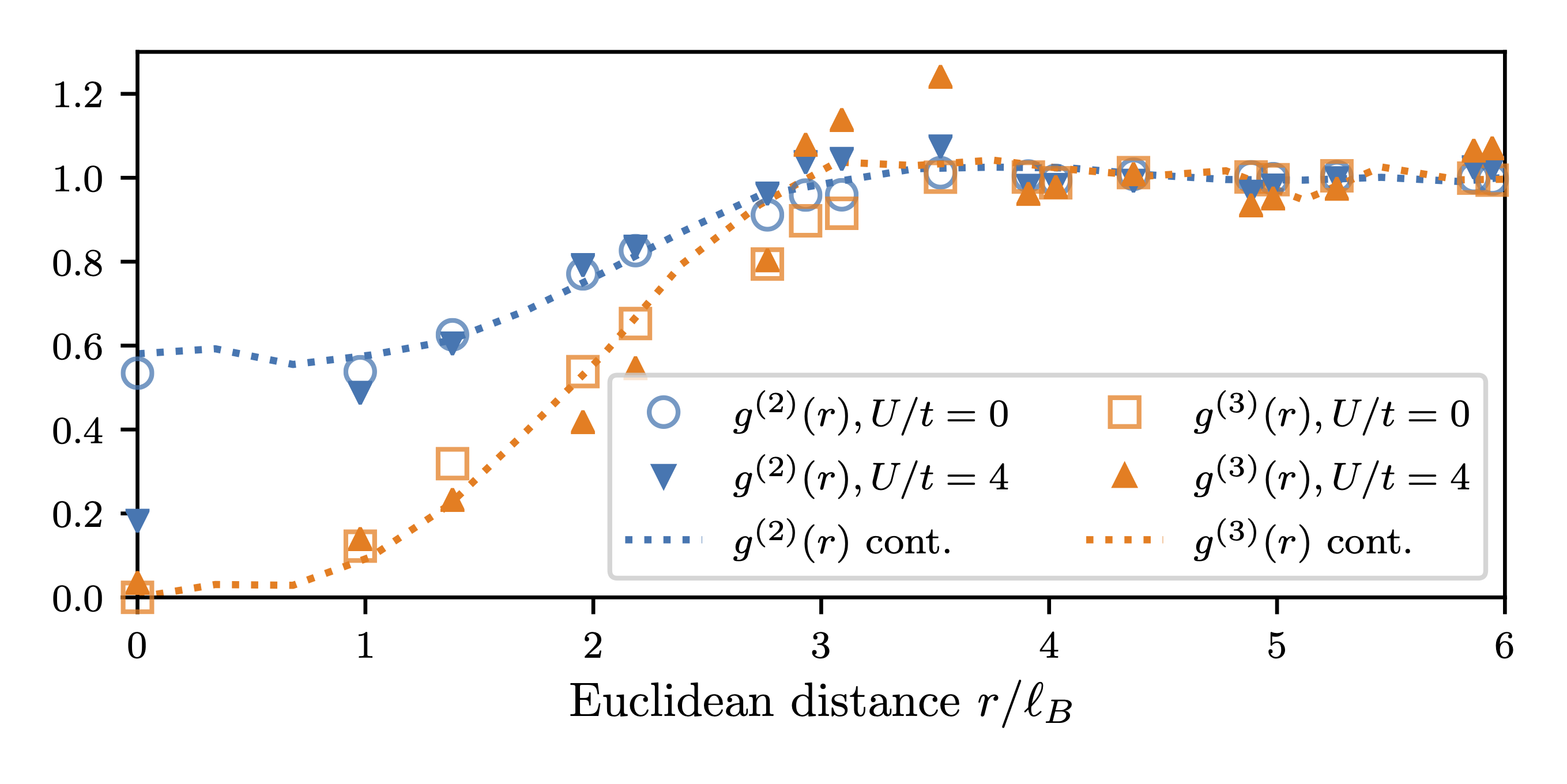}
	\caption{\label{fig:Correlations}Two- and three-particle correlations as function of the Euclidean distance in units of the magnetic length $\ell_B = a/\sqrt{2\pi\alpha}$ (same parameters as in Fig.~\ref{fig:ScalingCDW}(a)). At short distances, the $g^{(2)}$-function stays finite whereas the $g^{(3)}$-function drops to zero as expected for the continuum Pfaffian (dotted lines). For $U/t=0$ and $4$ we used $N_{\rm max}=2$ and $3$, respectively.}
\end{figure}

\textit{Two- and three-particle correlations.---} One of the key signatures of the continuum Pfaffian is its paired nature. In 2D, the exact Pfaffian (dotted lines in Fig.~\ref{fig:Correlations}) shows a strong suppression of $g^{(3)}(r \to 0)$ while $g^{(2)}(0) \neq 0$. This reflects the pairing of the bosons, which is built into the trial wave function of the Pfaffian.

Using DMRG we calculate the ground state of the three-body parent Hamiltonian $\hat{\mathcal{H}}^{(3)}$ in Eq.~\eqref{Eq:parentHamiltonian} on the lattice by setting $U/t = 0, N_{\rm max}=2$. Determining the two- and three-particle correlation functions,
\begin{align}
g_{\mathbf{i},\mathbf{j}}^{(2)}  &= \frac{\left\langle :\hat{n}_{\mathbf{i}}\hat{n}_{\mathbf{j}}:\right\rangle}{\left\langle\hat{n}_{\mathbf{i}}\right\rangle\left\langle\hat{n}_{\mathbf{j}}\right\rangle} = \frac{\left\langle \hat{n}_{\mathbf{i}}\hat{n}_{\mathbf{j}} \right\rangle}{\left\langle\hat{n}_{\mathbf{i}}\right\rangle\left\langle\hat{n}_{\mathbf{j}}\right\rangle} - \frac{\delta_{\mathbf{i,j}}}{\left\langle\hat{n}_{\mathbf{i}}\right\rangle},\\
g^{(3)}_{\mathbf{i}, \mathbf{j}} &= \left\langle :\hat{n}_{\mathbf{i}}\hat{n}_{\mathbf{i}}\hat{n}_{\mathbf{j}}:\right\rangle/(\left\langle\hat{n}_{\mathbf{i}}\right\rangle^2\left\langle\hat{n}_{\mathbf{j}}\right\rangle)\nonumber
\\
&= \frac{\left\langle \hat{n}_{\mathbf{i}}\hat{n}_{\mathbf{i}}\hat{n}_{\mathbf{j}}\right\rangle - \left\langle \hat{n}_{\mathbf{i}}\hat{n}_{\mathbf{j}}\right\rangle - 2\delta_{\mathbf{i,j}} \left\langle\hat{n}_{\mathbf{i}}\hat{n}_{\mathbf{i}}\right\rangle +2\delta_{\mathbf{i,j}} \left\langle\hat{n}_{\mathbf{i}}\right\rangle}{\left\langle\hat{n}_{\mathbf{i}}\right\rangle^2\left\langle\hat{n}_{\mathbf{j}}\right\rangle}.
\end{align}
with respect to a fixed test site $\mathbf{i}$ for varying site $\mathbf{j}$ we plot $g^{(2)}(r)$ and $g^{(3)}(r)$ as functions of the (Euclidean) distance $r = |\mathbf{i}-\mathbf{j}|$. We average over all available test sites in the bulk region of the cylinder. The close agreement of the parent Hamiltonian's ground state (open symbols in Fig.~\ref{fig:Correlations}) with the continuum result suggests that pairing is also present in this state on the lattice.

Finally, the experimentally relevant state $U/t = 4, N_{\rm max}=3$ (solid triangles in Fig.~\ref{fig:Correlations}) shows qualitatively similar correlations, in particular for $g^{(3)}(r)$. Only the on-site $g^{(2)}(0)$ is modified, indicating that the microscopic structure of the bosonic bound state has changed slightly on the smallest length scales.

\textit{Adiabatic state preparation.---}
\begin{figure}
	\includegraphics[width=\linewidth]{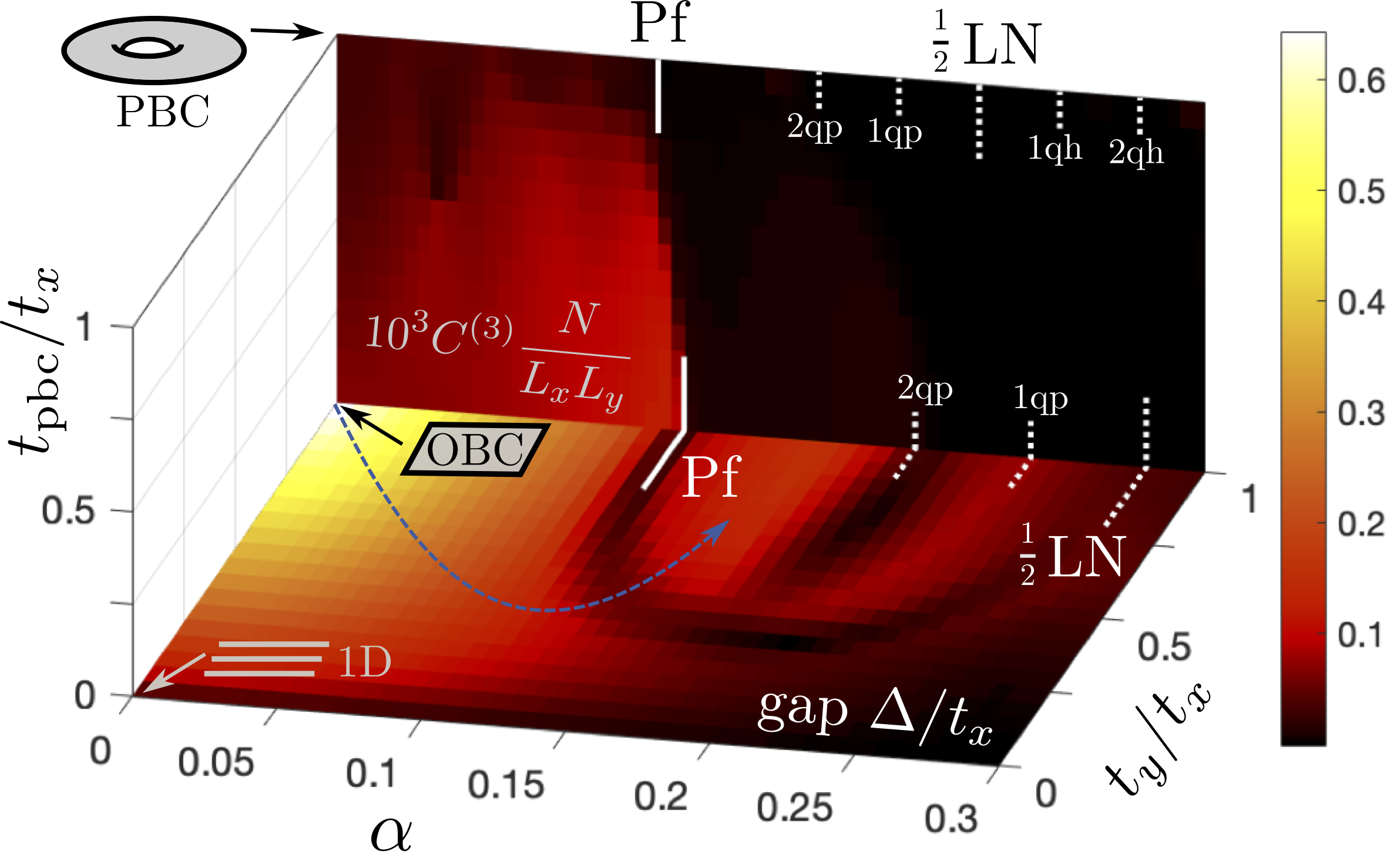}
	\caption{\label{fig:ED}Excitation gap $\Delta$ (plane) and $10^3\ C^{(3)}\ N/L_x L_y$ (wall) for various parameters as obtained by exact diagonalization of the HBH Hamiltonian for $L_x = L_y = 6, N=4, U/t_x = 4$. The blue dotted line indicates a possible path we propose for adiabatic preparation of the ground state, starting from a trivial superfluid state. We tune between 1D chains, open (OBC) and periodic (PBC) boundary conditions. `Pf' and `$\frac{1}{2}$LN' denote the Pfaffian and the Laughlin state respectively, `qp' and `qh' refer to quasi-particle and quasi-hole excitations of the Laughlin state  (see~\cite{supp} for details).}
\end{figure}
We performed an exact diagonalization (ED) of the HBH Hamiltonian for small, experimentally accessible systems~\cite{tai_17} (\mbox{$L_x=L_y=6, U/t_x = 4, N=N_{\rm max}=4$}) using various boundary conditions and anisotropic hopping along the $y$-direction. The key signatures of the Pfaffian discussed above are also visible in these systems as can be seen in Fig.~\ref{fig:ED}. In the 2D regime we find a finite-size excitation gap closely related to the charge gap observed in our DMRG calculations. Furthermore, the sharp drop of $C^{(3)}$ as a function of $\alpha$ or, equivalently, the filling factor $\nu=N/[\alpha L_x L_y]$ both on a plane and a torus shows the expected suppression of three-body correlations in the lattice system.

Various adiabatic preparation schemes for other FQH states have been proposed in recent years~\cite{popp_04, barkeshli_15, motruk_17, he_17, palm_20, andrade_20}. Making use of the large excitation gap found in ED a path like the one proposed by He et al.~\cite{he_17} (indicated by the blue line in Fig.~\ref{fig:ED}) provides a promising candidate. Another possibility could be the preparation of the Pfaffian starting from the closely related CDW and slowly turning on 2D couplings~\cite{motruk_17}.

\textit{Conclusions.---} We have found a close connection between the Pfaffian trial wave function in the continuum and the ground state of the Hofstadter-Bose-Hubbard model at filling $\nu = 1$. The most striking feature of the Pfaffian-like state is the associated suppression of on-site three-body correlations. We emphasize that two-body interactions are sufficient to realize the Pfaffian in currently accessible ultracold atom settings. We have proposed a realistic preparation scheme for small systems, and the closely related CDW in the quasi-1D-limit may provide a way to adiabatically prepare larger ground states. State-of-the-art techniques~\cite{bakr_09, sherson_10, preiss_15, bergschneider_18} allow for measurements of $n$-particle correlation functions in cold atom experiments providing a direct insight into the correlated nature of a state. Our work paves the way for future studies of excitations or the Hall response~\cite{repellin_19, motruk_20, repellin_20, buser_21} in the Pfaffian.

\begin{acknowledgments}
\ 

The authors would like to thank Frank Pollmann, Adam Kaufman, Markus Greiner, and Nathan Goldman for useful discussions. 
We acknowledge funding by the Deutsche Forschungsgemeinschaft (DFG, German Research Foundation) under Germany's Excellence Strategy -- EXC-2111 -- 390814868, and via Research Unit FOR 2414 under project number 277974659.
\end{acknowledgments}

\end{document}

\title{Supplemental Material: Bosonic Pfaffian State in the Hofstadter-Bose-Hubbard Model}

\author{F. A. Palm}
\affiliation{Department of Physics and Arnold Sommerfeld Center for Theoretical Physics (ASC), Ludwig-Maximilians-Universit\"at M\"unchen, Theresienstr. 37, D-80333 M\"unchen, Germany}
\affiliation{Munich Center for Quantum Science and Technology (MCQST), Schellingstr. 4, D-80799 M\"unchen, Germany}

\author{M. Buser}
\affiliation{Department of Physics and Arnold Sommerfeld Center for Theoretical Physics (ASC), Ludwig-Maximilians-Universit\"at M\"unchen, Theresienstr. 37, D-80333 M\"unchen, Germany}
\affiliation{Munich Center for Quantum Science and Technology (MCQST), Schellingstr. 4, D-80799 M\"unchen, Germany}

\author{J. L\'eonard}
\affiliation{Department of Physics, Harvard University, Cambridge, Massachusetts 02138, USA}

\author{M. Aidelsburger}
\affiliation{Munich Center for Quantum Science and Technology (MCQST), Schellingstr. 4, D-80799 M\"unchen, Germany}
\affiliation{Department of Physics, Ludwig-Maximilians-Universit\"at M\"unchen, Schellingstr. 4, D-80799 M\"unchen, Germany}

\author{U. Schollw\"ock}
\affiliation{Department of Physics and Arnold Sommerfeld Center for Theoretical Physics (ASC), Ludwig-Maximilians-Universit\"at M\"unchen, Theresienstr. 37, D-80333 M\"unchen, Germany}
\affiliation{Munich Center for Quantum Science and Technology (MCQST), Schellingstr. 4, D-80799 M\"unchen, Germany}

\author{F. Grusdt}
\affiliation{Department of Physics and Arnold Sommerfeld Center for Theoretical Physics (ASC), Ludwig-Maximilians-Universit\"at M\"unchen, Theresienstr. 37, D-80333 M\"unchen, Germany}
\affiliation{Munich Center for Quantum Science and Technology (MCQST), Schellingstr. 4, D-80799 M\"unchen, Germany}

\date{\today}

\maketitle

\onecolumngrid

\setcounter{equation}{0}
\setcounter{figure}{0}
\setcounter{table}{0}
\setcounter{page}{1}
\makeatletter
\renewcommand{\theequation}{S\arabic{equation}}
\renewcommand{\thefigure}{S\arabic{figure}}
\renewcommand{\bibnumfmt}[1]{[S#1]}

\subsection{A.$\quad$Method}
Our main results are obtained using the single-site variant of the density-matrix renormalization-group method~\cite{white_92, schollwoeck_11} with subspace expansion~\cite{hubig_15}. Exploiting the $\mathrm{U}(1)$ symmetry associated with the particle-number conservation, we determine the canonical ground state for fixed particle number $N$. We ensure convergence of all simulations by comparing the energy expectation value $\langle\hat{\mathcal{H}}\rangle$, the corresponding variance $\langle\hat{\mathcal{H}}^2\rangle-\langle\hat{\mathcal{H}}\rangle^2$, the local particle density $\langle\hat{n}_{x,y}\rangle$, and the local currents,
\begin{align}
\langle\hat{j}_{x,y}^{x}\rangle &= it\langle\hat{a}_{x+1,y}^{\dagger}\hat{a}_{x,y}\rangle + \mathrm{H.c.},\\
\langle\hat{j}_{x,y}^{y}\rangle &= it\mathrm{e}^{2\pi i\alpha x}\langle\hat{a}_{x,y+1}^{\dagger}\hat{a}_{x,y}\rangle + \mathrm{H.c.},
\end{align}
for different bond dimensions up to $\chi = 3000$.

\subsection{B.$\quad$Finite-size effects}
In this letter, we limit our studies to systems of at most $L_x \times L_y = 5 \times 49 = 245$ sites. Therefore, the results presented here are subject to finite-size effects. To exclude that the numerical results are significantly affected by the finite system size, we extrapolated the charge gap and the CDW order parameter to the limit $L_x \to \infty$ and find that the characteristic behavior persists in this limit.

\begin{figure}[hb]
	\includegraphics[width=\linewidth]{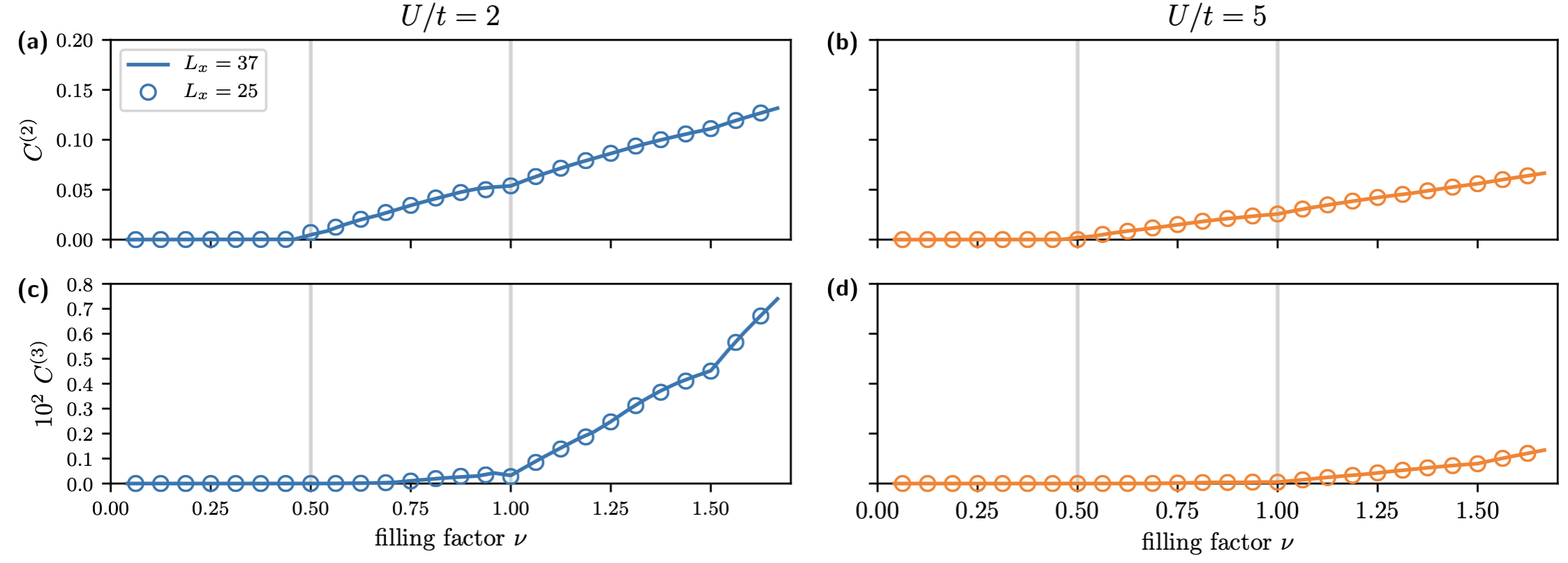}
	\caption{\label{fig:suppOnSiteCorr} Two- and three-particle on-site correlations for $U/t=2$ [(a) and (c)] and $U/t =5$ [(b) and (d)].}
\end{figure}

In Fig.~\ref{fig:suppOnSiteCorr} we determined the two- and three-particle on-site correlations per particle for $L_y=4, L_x=25$ and $L_x = 37$ to study the effect of the finite system size. We observe, that for both system sizes the results are in excellent agreement.

All our results, in particular the qualitative features discussed in this letter, are robust with respect to the system size and the remaining finite-size effects are quantitatively small.

%\subsection{C.$\quad$Particle number cut-off $N_{\rm max}=2$}
%Our results for the charge gap and the CDW order parameter were obtained for at most $N_{\rm max}=2$ bosons per site. This cut-off can be motivated by the fact that the three-particle on-site correlations at filling $\nu = 1$ are strongly suppressed and by the dilute systems ($N \ll L_x L_y$) studied here. Therefore, we believe this approximation captures the essential physics. To check its validity we performed DMRG calculations allowing for $N_{\rm max}=3$ bosons per site for some parameters (see Fig.~\ref{fig:Nmax3_supp}) and see that the qualitative behavior is the same. For $U/t=5$ even the quantitative corrections due to the larger cut-off $N_{\rm max}=3$ are small as expected for strong on-site repulsion.
%\begin{figure}[htb]
%	\includegraphics[width=\linewidth]{FigS2.png}
%	\caption{\label{fig:Nmax3_supp}(a) Charge gap for $N_{\rm max}=2$ (colored symbols) and $3$ (black and gray symbols). (b) CDW order parameter for $N_{\rm max}=2$ (blue, orange) and $3$ (purple, red). We see that the qualitative behavior is the same and already $N_{\rm max}=2$ captures the essential physics of the model. For $U/t=5$ we also find a close quantitative agreement.}
%\end{figure}

\subsection{C.$\quad$Grand-canonical ground state and charge gap}
\begin{figure}[htbp]
	\includegraphics[width=0.3\linewidth]{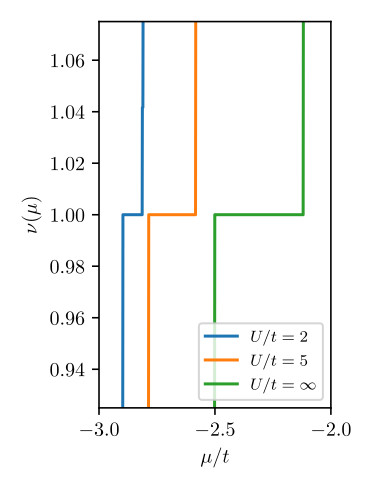}
	\caption{\label{fig:NuOfMu}Filling factor $\nu$ of the grand-canonical ground state as function of the chemical potential $\mu$ for $L_y = 4, L_x = 37$.}
\end{figure}

In order to find the grand-canonical ground state of the Hofstadter-Bose-Hubbard model, we start from the canonical ground state for a fixed particle number $N$, which has energy $E_{N}$. Using the chemical potential $\mu$, the grand-canonical ground state energy is given by
\begin{equation}
E^{\rm grand}(\mu) = \min_{N} \left(E_{N} - \mu N\right).
\end{equation}
From this we determine the particle number, or equivalently the filling factor $\nu$, of the grand-canonical ground state as a function of the chemical potential $\mu$ (see Fig.~\ref{fig:NuOfMu}), allowing for a very precise calculation of the charge gap. Therefore, we estimate the error of the charge gap at $\nu = 1$ by the variance of the Hamiltonian, $\mathrm{var}(\hat{\mathcal{H}}) = \langle \hat{\mathcal{H}}^2 \rangle - \langle \hat{\mathcal{H}}\rangle^2$, for the ground state at $\nu = 1$. We also use the variance to check the convergence of our calculations, and this error is negligibly small ($\mathrm{var}(\hat{\mathcal{H}}) \lesssim 10^{-5} t^2$) compared to the size of the charge gap. The linear extrapolation to $L_x \to \infty$ gives rise to an additional error from the least-square fit which is the dominant contribution to the overall uncertainty of the extrapolated charge gap.

\subsection{D.$\quad$Charge density wave on thin cylinders}
Previous studies of the continuum Pfaffian on thin cylinders showed that this state is closely related to the so-called Tao-Thouless states~\cite{bergholtz_06, seidel_06}. In particular, there is a state which consists of an alternation of doubly-occupied and unoccupied Landau level orbitals, $\ket{\hdots 2020\hdots}$. Assuming an infinite cylinder ($L_x = \infty$) and using the Landau gauge, $\vec{A} = xB\hat{\vec{y}}$, the wave functions describing Landau level orbitals take the form
\begin{equation}
\psi_{k,m}(x,y) = \mathrm{e}^{iky}\mathrm{e}^{-(x+k\ell_B)^2/2\ell_B^2}\ \mathrm{H}_m(x + k\ell_B^2),
\end{equation}
where $k$ is the eigenvalue of the momentum in $y$-direction, $\mathrm{H}_m$ are the Hermite polynomials and $\ell_B = \sqrt{1/B}$ is the magnetic length, where we used natural units where $\hbar = c = e = 1$. Here, $m$ denotes the Landau level, where $m=0$ is the lowest Landau level we consider here. These eigenstates are extended around the cylinder and localized in $x$-direction around $-k\ell_B^2$, where $k$ is quantized as $k_n = \frac{2\pi}{L_y a}n, n\in\mathbb{Z}$ with $L_y a$ the circumference of the cylinder.

\begin{figure}[htbp]
	\includegraphics[width=\linewidth]{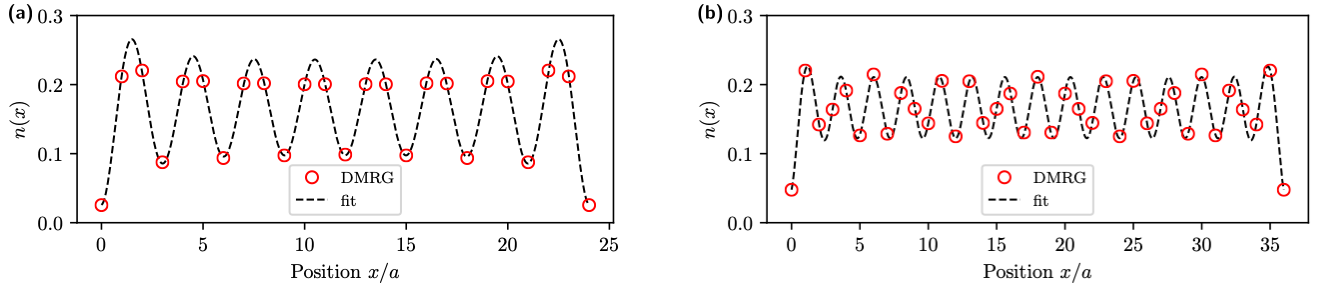}
	\caption{\label{fig:CDWprofile}Density profile and CDW fit for (a) $L_x=25, L_y = 4, U/t=2$ and (b) $L_x=37, L_y = 5, U/t=5$. The wavevectors of the fitted curve are $k_{\rm CDW} = 2.1018(9) a^{-1}, 2.6203(9) a^{-1}$, respectively, while the predicted values are $k_{\rm CDW}^{\rm pred} = 2.0944 a^{-1}$  and $2.6180 a^{-1}$, respectively.}
\end{figure}

Relating the magnetic length and the magnetic flux per plaquette via
\begin{equation}
\ell_B^2 = \frac{1}{2\pi\alpha}a^2,
\end{equation}
we can determine the centers of the Landau level orbitals to be at
\begin{equation}
x_{n} = -\ell_B^2 k_n = -\frac{1}{2\pi\alpha}a^2 \frac{2\pi}{L_y a}\cdot n = -\frac{n}{\alpha L_y}a, \quad n\in\mathbb{Z}.
\end{equation}
In the Tao-Thouless states $\ket{\hdots2020\hdots}$ every second orbital is doubly occupied so that the corresponding CDW has wavelength $\lambda_{\rm CDW} = |x_{2} - x_{0}| = 2a/\alpha L_y$, or equivalently wavevector
\begin{equation}
k_{\rm CDW} = \frac{2\pi}{\lambda_{\rm CDW}} = \frac{\pi \alpha L_y}{a}.
\end{equation}
The values of the wavevector obtained by a fit of the CDW profile
\begin{equation}
n(x) = n_0 + n_1\sin(k_{\rm CDW}x+\phi_0)\left(1+\frac{A}{2}\left(\mathrm{e}^{-x/\xi}+\mathrm{e}^{-(L_x-1-x)/\xi}\right)\right)
\end{equation}
to the numerical results as shown in Fig.~\ref{fig:CDWprofile} are in agreement with the expected wavevectors for the Tao-Thouless state.

\subsection{E.$\quad$Exact diagonalization and boundary conditions}
\begin{figure}[hbtp]
	\includegraphics[width=\linewidth]{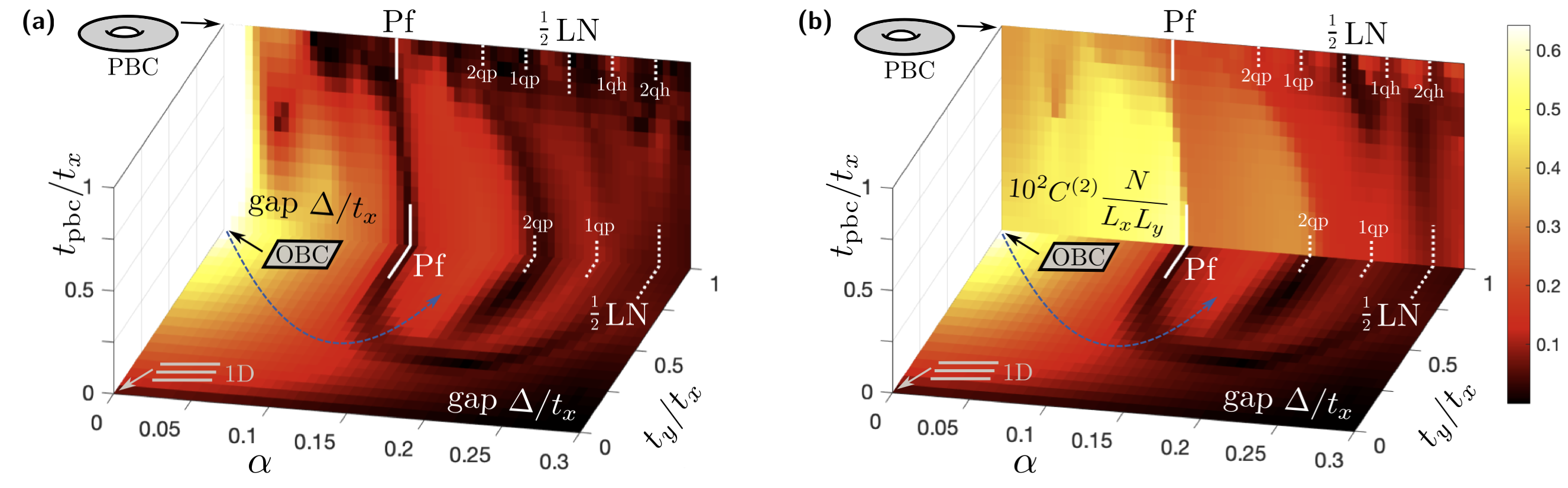}
	\caption{\label{fig:ED-supp}(a) Excitation gap $\Delta$ and (b) two-particle on-site correlations $10^2\ C^{(2)} N/L_x L_y$ for various parameters as obtained by exact diagonalization.}
\end{figure}

For small systems ($L_x = L_y = 6, N= 4$) exact diagonalization (ED) is a suitable method to study not only the ground state but also the excitation gap. In order to propose a path for the adiabatic preparation of the ground state at filling factor $\nu = 1$ we studied the system for 1D chains with tunable interchain hopping $t_y$, a 2D plane with open boundary conditions and tunable hopping $t_{\rm pbc}$ across the boundary, ultimately resulting in a torus for $t_{\rm pbc} = t_y = t_x$. In Fig.~\ref{fig:ED-supp}(a) we show the excitation gap and the two-particle on-site correlation $C^{(2)}$ for different setups. A possible path for adiabatic state preparation is indicated by the blue dotted line, where the minimal gap is of the order $\sim 0.05 t_x$.

The exact filling factor and hence the flux per plaquette at which we expect the Pfaffian and the Laughlin state depend on the choice of boundary conditions. For periodic boundary conditions the flux per plaquette is given by $\alpha_{\rm pbc} = N/[\nu L_x L_y]$, while for open boundary conditions we have $\alpha_{\rm obc} = N/[\nu (L_x-1) (L_y-1)]$. Thus, we expect the $\nu = 1$ Pfaffian ($\nu = 1/2$ Laughlin) state for periodic boundary conditions at $\alpha_{\rm Pf, pbc} = N/[L_x L_y]$ and $\alpha_{\rm LN, pbc} = 2N/[L_x L_y]$, respectively. These values are indicated by white lines in Fig.~\ref{fig:ED-supp}(b). In addition we included the one- and two-quasi-particle and quasi-hole excitations of the Laughlin state, obtained by removing/adding one or two flux quanta, respectively. We observe signatures of these states in both the excitation gap and the two-particle on-site correlation.

For open boundary conditions, the filling factors of the Pfaffian and the Laughlin state are shifted to $\nu_{\rm Pf} = N/[N-1]$ and $\nu_{\rm LN}=N/[2N-1]$, respectively, as can be derived from their exact wave functions in a disk geometry. Again, we indicated the corresponding values for $\alpha$ by white lines. From the ED results for the gap, Fig.~\ref{fig:ED-supp}(a), we conclude that the states on the torus and in the plane can be continuously connected.

\bibliographystyle{apsrev4-2}